# Linear Schrödinger equation with temporal evolution for front induced transitions

**MAHMOUD A. GAAFAR[1,2,*], HAGEN RENNER[1], ALEXANDER Y. PETROV[1,3] AND MANFRED EICH[1,4]**

[1]*Institute of Optical and Electronic Materials, Hamburg University of Technology, Eissendorfer Str. 38, 21073 Hamburg, Germany.*
[2]*Department of Physics, Faculty of Science, Menoufia University, 32511 Menoufia, Egypt*
[3] *ITMO University, 49 Kronverkskii Ave., 197101, St. Petersburg, Russia*
[4] *Institute of Materials Research, Helmholtz-Zentrum Geesthacht, Max-Planck-Strasse 1, Geesthacht, D-21502, Germany*
*\*Mahmoud.gaafar@tuhh.de*

**Abstract:** The nonlinear Schrödinger equation based on slowly varying approximation is usually applied to describe the pulse propagation in nonlinear waveguides. However, for the case of the front induced transitions (FITs), the pump effect is well described by the dielectric constant perturbation in space and time. Thus, a linear Schrödinger equation can be used. Also, in waveguides with weak dispersion the spatial evolution of the pulse temporal profile is usually tracked. Such a formulation becomes impossible for optical systems for which the group index or higher dispersion terms diverge as is the case near the band edge of photonic crystals. For the description of FITs in such systems a linear Schrödinger equation can be used where temporal evolution of the pulse spatial profile is tracked instead of tracking the spatial evolution. This representation provides the same descriptive power and can easily deal with zero group velocities. Furthermore, the Schrödinger equation with temporal evolution can describe signal pulse reflection from both static and counter-propagating fronts, in contrast to the Schrödinger equation with spatial evolution which is bound to forward propagation. Here, we discuss the two approaches and demonstrate the applicability of the spatial evolution for the system close to the band edge where the group velocity vanishes by simulating intraband indirect photonic transitions.

## 1. Introduction

In recent years, several theoretical predictions and experimental demonstrations were presented where the light propagating in guiding media is manipulated by a moving refractive index front [1–5]. The refractive index front is produced by a co/counter-propagating switching pulse via either Kerr nonlinearity [1,6–9] or free carrier injection [3–5]. In this case, the front induces an indirect transition of the optical state between two points in the dispersion diagram, thus, a change both in frequency and wavenumber. The dispersion engineering of the guided modes allows manipulation of light to an extent that goes far beyond the simple Doppler frequency shift in free space. Concepts of front induced transitions (FITs) are proposed and realized for frequency manipulation [6,8–10], light stopping and optical delays [11,12], bandwidth/time duration manipulation [13,14] as well as optical isolation [15]. For the description of pulse propagation in a nonlinear medium usually a slowly varying approximation is used which leads to a nonlinear Schrödinger equation [16–19]. For the case of the FIT the pump effect can be approximated by the dielectric constant $\Delta\varepsilon(r,t)$ perturbation in space and time [19]. Thus, a linear Schrödinger equation (LSE) can be used. At the same time, in waveguides with weak dispersion usually the spatial evolution of the pulse temporal profile is tracked [6,8,19]. This is a natural choice for the comparison with experimental data as detectors are usually measuring temporal pulse profiles at certain positions in the optical signal line. As we show in this publication such a choice becomes problematic for optical systems with large dispersion, where group velocity dramatically changes with frequency, or even impossible when group velocity vanishes. For the description of FITs in such systems a linear Schrödinger equation can be used where the temporal evolution of the pulse spatial profile is tracked instead of tracking the spatial evolution. This representation can easily deal with zero group velocities, as, for example, is the case at the band edge of photonic crystals. Finally, the simulation results can be also compared to experimental results by converting the spatial profile of the pulse at a time when the signal stopped to interact with the front into a temporal profile at any defined position. In this article we discuss the temporal and spatial evolution and demonstrate the applicability of the spatial evolution for the system close to the band edge by simulating intraband indirect FITs in a system with hyperbolic dispersion as an example. The hyperbolic dispersion is a good approximation for the dispersion of a weak Bragg grating in a dispersionless waveguide [20].

A perturbation of the dielectric constant $\Delta\varepsilon(t,z)$ in space and time also leads to a perturbation in space and time of the local dispersion relation. The dispersion relation can be described as a function of wavenumber versus frequency or vice versa. Two simplified situations of this perturbation can be considered: either a frequency independent horizontal shift $\Delta\beta_D(t,z)$ (Fig. 1(a)), or a wavenumber independent vertical shift $\Delta\omega_D(t,z)$ (Fig. 1(b)) of the dispersion relation. Here, the solid curves represent the dispersion bands of the original (unperturbed) modes, while the dashed curves indicate the switched (perturbed) states due to the presence of the perturbation after the front has passed a position. We consider the signal propagation in the dispersive waveguide with envelope function $A(t,z)$ for the carrier angular frequency $\omega_0$ and carrier wave number $\beta_0$. In the following discussions the angular frequency $\omega$ and wavenumber $\beta$ are defined in respect to the carrier frequency and carrier wavenumber correspondingly, so that the total angular frequency is $\omega_0 + \omega$ and the total wavenumber are $\beta_0 + \beta$.

We start with a derivation of the equation for the spatial evolution of the temporal profile without front. The phase accumulated with propagation in the dispersive waveguide is frequency dependent. Thus we consider the envelope function in the temporal frequency domain as $A(\omega,z)$ and write the propagation equation in a waveguide without perturbation as:

$$\frac{\partial A(\omega,z)}{\partial z} = i\beta(\omega)A(\omega,z) \quad (1)$$

This equation can be converted into a time domain equation. To obtain a differential equation a Taylor expansion of the dispersion relation is used as:

$$\beta(\omega) = \sum_{n=1}^{N} \frac{\beta_n}{n!} \omega^n \quad (2)$$

Where $\beta_n = \partial^n \beta / \partial \omega^n$ at $\omega = 0$. The first coefficient $\beta_1$ corresponds to the inverse group velocity $v_{g0}^{-1}$ of the carrier wave. The Fourier transform leads to the linear Schrödinger equation (LSE) with spatial evolution of $A(t,z)$ [21]:

$$\frac{\partial A(t,z)}{\partial z} = \sum_{n=1}^{N} i^{(n-1)} \frac{\beta_n}{n!} \frac{\partial^n A}{\partial t^n} \quad (3)$$

The front induced perturbation of the propagation constant can be now added to equation (3). When we integrate over z-direction, it will lead to an additional phase shift proportional to $\Delta\beta_D(t,z)$ and accordingly the equation can be written as:

$$\frac{\partial A(t,z)}{\partial z} = \sum_{n=1}^{N} i^{(n-1)} \frac{\beta_n}{n!} \frac{\partial^n A}{\partial t^n} + i\Delta\beta_D(t,z)A \quad (4)$$

An additional assumption can be made of a front propagating with constant velocity $v_f$ and not changing during the propagation $\Delta\beta_D(t,z) = \Delta\beta_D(t - z/v_f)$. In this case a new time parameter can be introduced $t' = t - z/v_f$ in the equation (4). This leads to the following equation:

$$\frac{\partial A(t',z)}{\partial z} = \left(\frac{1}{v_{g0}} - \frac{1}{v_f}\right)\frac{\partial A}{\partial t'} + \sum_{n=2}^{N} i^{(n-1)} \frac{\beta_n}{n!} \frac{\partial^n A}{\partial t'^n} + i\Delta\beta_D(t')A \quad (5)$$

In some systems the carrier frequency of the envelope can be chosen in such a way that $v_{g0} = v_f$. In this case the first term on the right side can vanish [21]. The obtained description moves the observer in the time frame of the moving front. Thus the signal position in time is now measured in respect to the front. The space parameter $z$ is not changed and thus corresponds to the original propagation length in the waveguide. In this frame the perturbation happens everywhere at the same time (simultaneously). Thus, the signal in this frame accumulates only a vertical frequency shift without a wavenumber shift [22]. The frequency shift obtained from the Fourier transform of the envelope $A(t',z)$ after propgation a distance $z = L$ is the same as accumulated by the signal in the stationary frame $A(t,z)$. Thus, the frequency shift can be directly obtained from this consideration. At the same time the wavenumber shift can be obtained only by transforming back into the stationary frame.

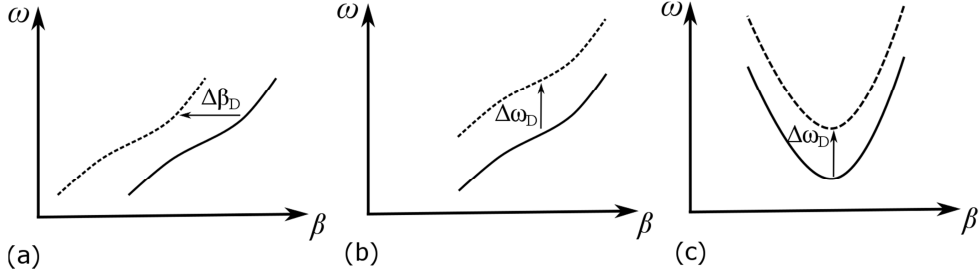

Figure 1: Schematics of different dispersion relations with perturbation. Here, the solid curves represent the dispersion bands of an original (unperturbed) mode, while the dashed curves indicate the switched (perturbed) state. In the systems with small dispersion, the shift can be approximated either (a) by a constant horizontal shift $\Delta\beta_D$, or (b) by a constant vertical shift $\Delta\omega_D$. (c) For systems with large dispersion, the vertical shift cannot be represented by a horizontal shift, instead.

This description of dispersion as a $\beta(\omega)$ works very well in many systems with small dispersion such as weakly guiding waveguides and fibers [6,8]. However, for some other systems, which, for example, have non-unique dispersion relations, the $\beta(\omega)$ description and the wavenumber shift approximation become inapplicable. E.g., such an equation cannot be applied to the pulse propagation close to the band gap. The $\beta(\omega)$ function is not unique there. Also at the band edge the dielectric constant perturbation leads mainly to a frequency shift of the band diagram and not to a wavenumber shift [23]. These non-unique dispersion functions frequently appear in periodic structures [23], such as photonic crystal waveguides [24], photonic crystal fibers [25], Bragg gratings [20,26,27]. A schematic example of such dispersion relation is shown in Fig. 1(c). Furthermore, the equation with spatial evolution is derived assuming an enevelope propagating only in the forward direction. No backward propagation in space is possible; therefore it is not valid also to describe signal pulse reflection even from a static perturbation (nonmoving front).

In nonlinear fiber optics an alternative description is known. Yulin et al. described the evolution of the field in the nonlinear photonic bandgap fiber in the spectral range at the cutoff frequency [28]. They used the time coordinate $t$ instead of the propagation coordinate $z$ as the evolution variable. Furthermore, the optical mode propagation close to the band gap of fiber Bragg gratings with cross phase modulation from a pump has been described by coupled mode equations in Ref [29] which also lead to a LSE with temporal evolution. In this paper we would like to present the application of a LSE with temporal envelope evolution to describe a front induced transition in dispersive waveguides. The results are compared to the predictions from the phase continuity criterion for complete transitions [5,10].

To introduce time evolution we now consider the dispersion relation in the form of $\omega(\beta)$ instead of $\beta(\omega)$. Thus, the phase accumulated by the signal with evolving time is a function of its spatial frequency or wavenumber:

$$\frac{\partial A(t,\beta)}{\partial t} = i\omega(\beta)A(t,\beta) \quad (6)$$

Following all the steps that lead to equation (5) in case of spatial evolution we now obtain:

$$\frac{\partial A(t,z')}{\partial t} = (v_{g0} - v_f)\frac{\partial A}{\partial z'} + \sum_{n=2}^{N} i^{(n-1)}\frac{\omega_n}{n!}\frac{\partial^n A}{\partial z'^n} + i\Delta\omega_D(z')A$$

(7)

In this equation, $\omega_n = \partial^n\omega/\partial\beta^n$ are the dispersion coefficients associated with the Taylor series expansion of the dispersion function $\omega(\beta)$ at $\beta = 0$, and $z' = z - v_f t$, where $v_f$ is the front group velocity. In this case the temporal evolution of the envelope is a function of the spatial dispersion and the frequency shift of the dispersion relation with relative position with respect to the front. Thus, the time now corresponds to the time in the stationary frame. The front is not moving in the considered frame $(t, z')$ and thus represents a stationary perturbation where the frequency of the signal is not changed. With the time parameter the wavenumber of the envelope is changing. This change is the same as for a signal in the stationary frame. The temporal profile of the signal can be obtained explicitly by observing the temporal evolution at a fixed position $z$, e.g., at the waveguide output.

The time evolution representation of the signal propagation is a more mechanistic approach as it tracks the spatial distribution of the signal pulse with time similar to the motion of particles. It is also closer to the time-dependent Schrödinger equation used in quantum mechanics. The spatial evolution representation is more common in optics as signals with certain time function are usually launched at the input of the waveguide and their time function at the input is measured. But such consideration, as discussed above, does not allow the description of systems with non-unique $\beta(\omega)$ function. As we show, the time evolution approach will allow to do that, it also can consider systems where the signal pulse is reflected and propagates in the backward direction or splits in forward and backward propagating waves. It also includes the reflection of the signal from a stationary perturbation. In the following we present several numerically integrated results for moving and static fronts and compare obtained results to theoretical predictions.

## 2. Simulation

### 2.1 Linear (unique) dispersion function

As a first example, we investigate a dispersion-free case where the unperturbed and perturbed dispersion functions are linear functions with equal slopes. This can be a good approximation for waveguides with weak dispersion and small band diagram shift. We are solving equation (7) by split-step Fourier method [21,30] where the additional phase due to dispersion is calculated in Fourier space and due to the perturbation in the real space. First, we simulated the signal pulse propagation in a system with linear (unique) dispersion relation given by

$\omega(\beta) = (c/2) \cdot \beta$, where $c$ is the speed of light. Figure 2(a) shows the dispersion relations for both unperturbed (solid curve) and perturbed –frequency upshifted- (dashed curve) states. In our simulation, the input signal pulse is centered at 200 THz and has a spatial width of 5 mm (temporal width $\approx$ 31 ps). The velocity of the front $v_f$ is equal to the group velocity of light $c$, while the group velocity of the signal $v_{g1}$ is equal to $c/2$. Red and blue dots indicate the initial and final states of the signal pulse on the band diagram, respectively. The orange line represents the phase continuity line with a slope equal to the group velocity of the front [3,5,13]. The final state of the signal is defined by the intersection point between the phase continuity line and the band diagram [3,4]. From Fig. 2(a), the expected wavenumber shift of the signal pulse after interaction with the front can be calculated using $\Delta\beta = \Delta\omega_{D_{max}}/(v_f - v_{g1})$, where $\Delta\omega_{D_{max}}$ is the maximum vertical band diagram shift in frequency [5]. For $\Delta\omega_{D_{max}} = 2.5$ THz, the expected wavenumber shift is $\Delta\beta = 1.6660 \cdot 10^4$ m$^{-1}$. We can also calculate the expected frequency change using the relation $\Delta\omega = v_f \Delta\beta$ [5], which yields 4.98 THz. In this simulation, the band diagram shift induced by the front is described by the function $\Delta\omega_D(z') = \Delta\omega_{D_{max}} \cdot [(1 + \tanh(z'/z_f))/2]$, where $z_f$ is the spatial front width. The total length of the simulated structure is 20 cm, the spatial front width is 3 mm, while the time step for the split-step Fourier method $\Delta t$ is chosen to be 0.2 ps such that the results using $\Delta t$ between 0.2-0.5 ps converges [21].

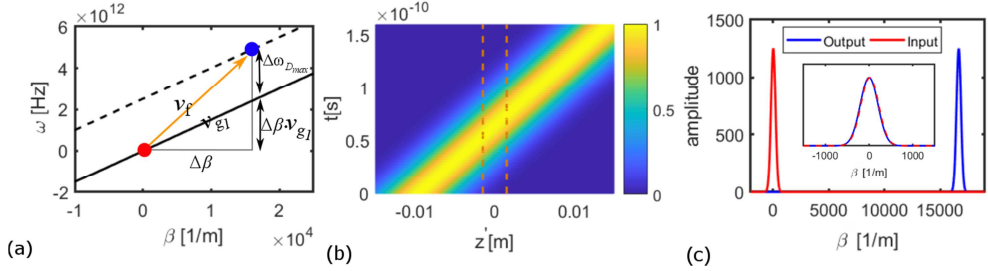

Figure 2: Simulation results of a signal pulse interacting with a refractive index front in a system with linear dispersion relation. (a) Dispersion relation. The solid black curve represents the dispersion band of an original (unperturbed) mode, while the vertically frequency shifted dispersion is indicated by the dashed curve. Initially, the group velocity of the signal pulse $v_{g1}$ and the velocity of the front $v_f$ are co-directed. $\Delta\omega_{D_{max}}$ is the maximum vertical band diagram shift in frequency. The frequency $\omega$ and wavenumber $\beta$ are the deviations from the carrier frequency $\omega_0$ and carrier wavenumber $\beta_0$ respectively. The orange line represents the phase continuity line with a slope equal to the velocity of the front. Red and blue dots indicate the initial and final states of the signal on the band diagram, respectively. The final state of the signal is defined by the intersection point between the phase continuity line and the band diagram, (b) Spatial dynamics of signal pulse along propagation. The dashed vertical orange lines represent the boundaries of the front's spatial width (3mm). The false color indicates the intensity of the electric field of the signal. (c) Input (red curve) and output (blue curve) wavenumbers of the signal pulse. In the inset we plot both signal spectra shifted to each other. Simulation parameters are given in the text.

The temporal dynamics of the signal pulse is shown in Fig. 2(b). As we expect from the band diagram, the signal pulse transmits through the front (centered at $z' = 0$ position, and its spatial width boundaries marked by vertical dashed orange lines) and undergoes an indirect interband photonic transition [3–5].

The spatial Fourier spectra of the input (red curve) and output (blue curve) signal pulses, respectively, are shown in Fig. 2(c). As we can see from the simulation, the wavenumber shift is equal to $\Delta\beta = 1.6659 \cdot 10^4$ m$^{-1}$, which accurately fits with the shift expected from the consideration made with respect to Fig. 2(a). The obtained shift is not dependent on the front function and depends only on the final perturbation of the dispersion relation. Furthermore, no broadening or narrowing of the output signal pulse is obtained (as shown in the inset, by plotting both signal spectra shifted to each other), as also expected for the identical linear dispersion relations of the unperturbed and perturbed bands.

### 2.2 Dispersion relation at the band edge

Next, we consider signal pulse propagation close to a band edge. This kind of dispersion relation, which is shown in Figs. 1(c) and 3, appears in periodic structures, such as photonic crystal waveguides [24], photonic crystal fibers [25], and Bragg gratings [26,27]. In this section, we will consider 3 different interaction scenarios: signal reflection from a co-propagating front (Fig. 3(a)), signal reflection from a counter-propagating front (Fig. 3(b)), and finally signal reflection from a static front (Fig. 3(c)). In all three cases we use the same hyperbolic dispersion relation $\omega + \omega_0 = \omega_{PBG} + \Delta\omega_{PBG} \cdot \sqrt{1 + [(\beta + \beta_0 - \beta_{PBG})^2/\Delta\beta_{PBG}^2]}$, with a half photonic band gap (PBG) opening of $\Delta\omega_{PBG} = 10$ THz and the upper band edge at $\omega_0 = 200$ THz and $\beta_0 = 1.33 \cdot 10^6$ m$^{-1}$. $\Delta\beta_{PBG} = \Delta\omega_{PBG}/v_{g\infty}$ is the parameter that is chosen in such a way, that away from the band edge the dispersion relation converges to a straight line with group velocity $v_{g\infty} = c/2$. Here, $\omega_{PBG}$ and $\beta_{PBG}$ are the center frequency and wavenumber of the PBG. For all three cases, the front spatial profile is described again by the function $\Delta\omega_D(z') = \Delta\omega_{D_{max}} \cdot [1 + \tanh(z'/z_f)/2]$. The total length of the simulated structure, spatial widths of the signal and of the front, as well as the time step $\Delta t$ are the same as in the linear dispersion case discussed above.

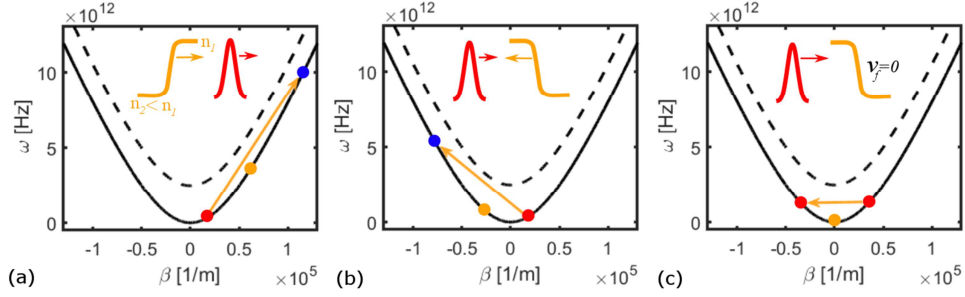

Figure 3: Interactions between signal pulse and index front. (a) Initial group velocities of the signal and of the front are co-directed. (b) Initial group velocities of the signal and of the front are counter-directed. (c) Signal interacting with a static front. Schematic representations of all scenarios are shown in the insets. The orange line represents the phase continuity line with a slope equal to the group velocity of the front. Again, red and blue dots indicate the initial and final states of the signal on the band diagram, respectively. The final state of the signal is defined by the intersection point between the phase continuity line and the band diagram. For the static front, the frequency of the signal does not change; therefore the final state of the signal is also marked by red color. Simulation parameters are given in the text.

#### 2.2.1 Signal and front are co-propagating

In this section we simulate signal pulse interaction with a co-propagating front. This scenario corresponds to Fig. 3(a). Schematic representation of this interaction is shown in the inset. Similar situation, but far away from band edge, has been already described by the LSE with spatial evolution [19,31,32]. In our simulation, the input signal pulse is also centered at 200 THz ($\beta = 1.366 \cdot 10^4$ m$^{-1}$) and has a spatial width of 5 mm, the velocity of the front $v_f$ is equal to $c/3.1$, and $\Delta\omega_{D_{max}} = 2.5$ THz. Under these conditions, the phase continuity line, which has a slope equal to the velocity of the front, can reach the shorter wavelength position in the same band without cutting into the shifted band, and therefore the signal undergoes an intraband transition [10,33]. This intraband transition manifests itself as a forward reflection from the co-propagating front. For such a dispersion relation, the expected wavenumber shift can be predicted from the crossing points between the phase continuity line and the unperturbed band diagram, as can be seen in Fig. 3(a). Here, the phase continuity line cuts the unperturbed band at $\beta = 1.1725 \cdot 10^5$ m$^{-1}$, and accordingly the expected wavenumber shift is $\Delta\beta = 1.0359 \cdot 10^5$ m$^{-1}$. We expect in this case the signal to be reflected from the front, thus to stay in the unperturbed medium.

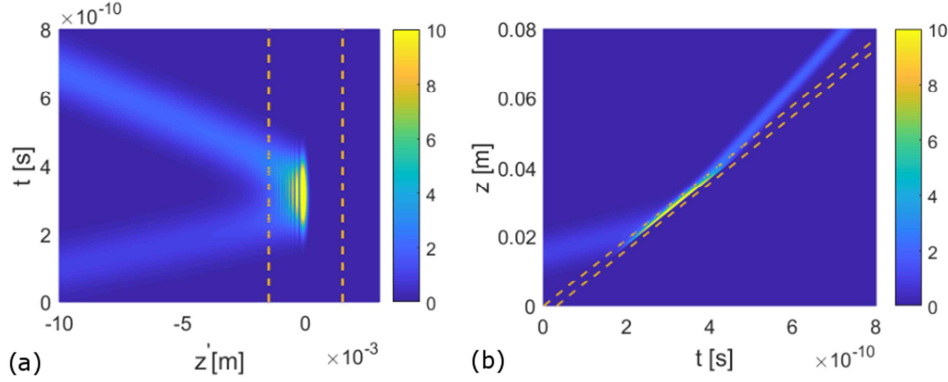

Figure 4: Simulation of signal pulse interaction with a co-propagating refractive index front, corresponding to the transition shown in Fig. 3(a). (a) Temporal evolution of the signal in the spatial frame of the front. Here, the dashed orange lines represent the boundaries of the front's spatial width (3mm). (b) Temporal evolution of the same signal represented in the stationary frame. Temporal pulse compression is seen after interaction with the front. The signal is reflected in the forward direction. The false color indicates the intensity of the electric field of the signal. Simulation parameters are given in the text.

The spatial dynamics of the signal pulse along propagation is shown in Fig. 4. The temporal evolution in Fig. 4(a) is strikingly analogous to the spatial evolution representation away from band edge as in Ref. [19] where the front is fixed in time as in eq (5). As we can see from Fig. 4(b), the signal pulse is reflected by the front experiencing a temporal and a spatial compression. The frequency width of the final state of the signal pulse is increasing. This can be explained schematically by projecting the initial frequency width, via the phase continuity lines, on the unperturbed dispersion relation [13]. From this geometrical consideration the temporal compression factor $t_{input}/t_{output}$ can be derived as $[(1 - v_f/v_{g1})/(v_f/v_{g2} - 1)]$, where $v_{g1}$ and $v_{g2}$ are the group velocities of the signal before and after the interaction with the front. The compression factor will increase when the front velocity will approach the slope of the dispersion relation away from the band edge $c/2$, provided that the phase continuity line does not cut through the perturbed dispersion function.

We achieved from the simulation, that the wavenumber shift is equal to $\Delta\beta = 1.0360 \cdot 10^5$ m$^{-1}$, which fits the expected calculated shift from the intersection between phase continuity line and band diagram $\Delta\beta = 1.0359 \cdot 10^5$ m$^{-1}$. Furthermore, we observe a wavenumber broadening of $\approx 2$ times, as expected from the shape of the dispersion relation.

### 2.2.2 Signal and front are counter-propagating

Next, the situation where the signal pulse and front are counter-directed is considered. This scenario corresponds to Fig. 3(b). A schematic representation of this interaction is also shown in the inset. Here, the signal pulse frequency and therefore the signal velocity is the same as for the co-propagating case (centered at 200 THz ($\beta = 1.366 \cdot 10^4$m$^{-1}$)), while the front velocity is decreased to $-c/5.3$, such that the phase continuity line does not cut the perturbed dispersion curve, and therefore again an intraband transition occurs (cf. Fig. 3(b)). The difference to the co-propagating case is that the signal pulse here will be reflected and move in the backward (opposite) direction compared to its initial propagation direction. Such a signal reflection from a counter-propagating front has also been simulated by finite-difference time-domain (FDTD) method in Ref. [34]. We have to mention that the Schrödinger equation with spatial evolution (eq. (5)) will be not able to simulate this case, as it allows the propagation only in the same spatial direction. [16–19,31,32].

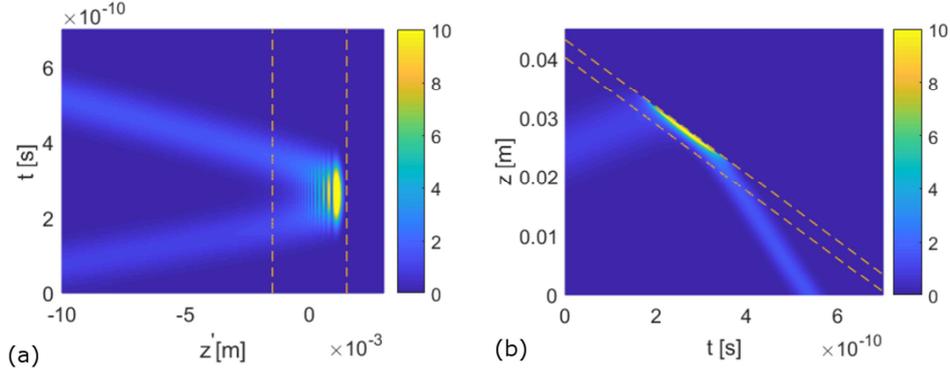

Figure 5: Simulation of a signal pulse interaction with a counter-propagating index front, corresponding to the transition shown in Fig. 2(b). (a) Temporal evolution of the signal in the spatial frame of the front, (b) Temporal evolution of the signal in the stationary frame. Signal reflected in the backward direction can be seen. The false color indicates the intensity of the electric field of the signal. Simulation parameters are given in the text.

The spatial dynamics of the signal pulse along propagation is shown in Fig. 5. The signal pulse is reflected from the counter-propagating front with a temporal compression. This is again connected to increasing spectral frequency width of the final state of the signal pulse. The compression factor is slightly decreased, compared to the co-propagating case. Again, this can be explained schematically by projecting the initial frequency width, via the phase continuity lines, on the unperturbed dispersion relation [13]. Here, the phase continuity line cuts the unperturbed band at $\beta = -7.8348 \cdot 10^4 \mathrm{m}^{-1}$, and, accordingly, the expected wavenumber shift is $\Delta\beta = 0.92008 \cdot 10^5$ m$^{-1}$, compared to $\Delta\beta = 0.92010 \cdot 10^5$ m$^{-1}$ achieved from the simulation (not presented here).

### 2.2.2 Interaction with a static front

Finally, we simulate the signal pulse reflection from a static front (Fig. 3(c)). Again as the LSE with spatial evolution (eq. 5) is valid only for forward propagation along the z-coordinate, it will not be able to describe the signal reflection from a static front, in contrast to the present LSE with temporal evolution (eq. 7) used here. The temporal evolution can as well describe the partial signal pulse transmission through a weak static front. In Fig. 3(c), the phase continuity line, which has a zero slope, does not cut the perturbed dispersion curve, and therefore reflection from the static front occurs (Fig. 2(c)). As the front is not moving, we only introduce a spatial perturbation in the structure; therefore the signal wavenumber will be altered without frequency shift. The temporal dynamics of the signal pulse along propagation is shown in Fig. 6. Naturally, there is no compression and $z$ and $z'$ represenation are identical in this case.

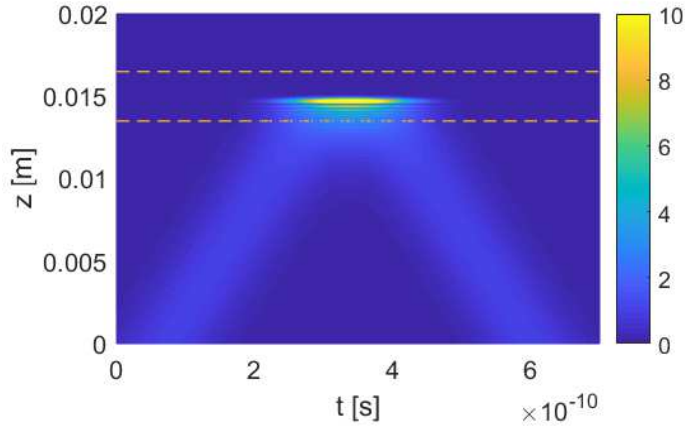

Figure 6. Temporal evolution of the signal in a waveguide with a static front, corresponding to the transition shown in Fig. 3(c). The dashed orange lines represent the boundaries of the spatial width of the static front (3mm).

## 3. Conclusions

We showed in this paper, that a linear Schrödinger equation formulation, where the spatial evolution of the pulse temporal profile is tracked becomes impossible for optical systems for which the group index or higher dispersion terms become infinite. Such a situation may occur, for instance, when the group velocity vanishes at the band edge of a photonic crystal. Therefore, for the description of front induced transitions in such systems a linear Schrödinger equation should be used where, instead, the temporal evolution of the pulse spatial profile is tracked. In this case, the dispersion relation is represented in the equation by spatial dispersion. This representation can easily deal with zero group velocities, as, for example, is the case at the band edge of photonic crystals. The integration of the equation allows to obtain the spatial and the temporal profile of the signal after the interaction with the front. Furthermore, the Schrödinger equation with spatial evolution tracks the signal propagation only in the forward direction. Therefore, it is not applicable for the description of the signal pulse reflection from both static and counter-propagating fronts, in contrast to the Schrödinger equation with temporal evolution presented here. We demonstrate the applicability of the temporal evolution representation for a system close to the band edge by simulating intraband front induced transitions in a system with hyperbolic dispersion.

## Acknowledgements

We would like to acknowledge the support of the German Research Foundation (DFG) under Grant No. 392102174. We further acknowledge the discussion with Andrei Sukhorukov.